\documentclass{article}

\usepackage{arxiv}

\usepackage[utf8]{inputenc} 
\usepackage[T1]{fontenc}    
\usepackage{hyperref}       
\usepackage{url}            
\usepackage{booktabs}       
\usepackage{amsfonts}       
\usepackage{nicefrac}       
\usepackage{microtype}      
\usepackage{lipsum}
\usepackage{graphicx}
\graphicspath{ {./images/} }

\usepackage{amsmath}
\usepackage{natbib}

\usepackage{siunitx}
\usepackage{moreverb}
\usepackage{mathtools, cuted}
\usepackage{rotating}

\title{Sources of low-frequency $\delta^{18}$O variability in coastal ice cores from Dronning Maud Land}

\author{
Stéphane Vannitsem \\ Royal Meteorological Institute of Belgium, Avenue Circulaire, 3, 1180 Brussels \\ 
\texttt{Stephane.Vannitsem@meteo.be} \\
\And
 David Docquier \\ Royal Meteorological Institute of Belgium, Avenue Circulaire, 3, 1180 Brussels \\
 \texttt{David.Docquier@meteo.be} \\
\And 
 Sarah Wauthy \\ Universit\'e Libre de Bruxelles, Brussels, Belgium \\
 \texttt{Sarah.Wauthy@ulb.be} \\
\And 
 Matthew Corkill \\ University of Tasmania, Hobart, Tasmania, Australia \\
 \texttt{matthew.corkill@utas.edu.au} \\
 \And 
 Jean-Louis Tison \\ Universit\'e Libre de Bruxelles, Brussels, Belgium \\
 \texttt{Jean-Louis.Tison@ulb.be} \\
}

\begin{document}

\maketitle

\begin{abstract}
The low-frequency variability of the $\delta^{18}$O recorded in ice cores (FK17 and TIR18) recently drilled at two different locations in Dronning Maud Land (Antarctica), is investigated using multi-taper spectral method and singular spectrum analysis. Multiple dominant peaks emerge in these records with periods between 3 and 20 years. The two sites show distinct spectral signatures, despite their relative proximity in space (about 100 km apart), suggesting that different processes are involved in generating the variability at these two sites. 
In order to clarify which processes are acting on  $\delta^{18}$O at these two locations, the impact of several climate indices as well as sea ice area is investigated using a causal method, known as the Liang-Kleeman rate of information transfer. The analysis of the origin of this low-frequency variability from external sources reveals that El Niño-Southern Oscillation (ENSO), the Pacific Decadal Oscillation (PDO), the Southern Annular Mode (SAM), the Dipole Mode Index (DMI) and the sea ice area display important causal influences on $\delta^{18}$O at FK17. For TIR18, the main influences are from ENSO, PDO, DMI, the sea ice area, and the Atlantic Multidecadal Oscillation (AMO), revealing the complexity of the interactions in Dronning Maud Land. The two locations share several drivers, but also show local specificities potentially linked to ocean proximity and differences in air mass trajectories. The implication of these findings on the low-frequency variability in the two ice cores is discussed. 

\end{abstract}

\section{Introduction}
\label{intro}

While considerable ice losses have been reported in Greenland, West Antarctica and the Antarctic Peninsula, East Antarctica remains close to a state of ice-sheet mass balance equilibrium, despite large uncertainties in mass balance measurements for this region \citep{Fox-Kemper2021,Otosaka2023}. Dronning Maud Land is located in East Antarctica between $\sim$~20$^\circ$W and $\sim$~45$^\circ$E and covers about a sixth of the Antarctic continent. This region has experienced a relatively strong ice mass gain between 2002 and 2019, which partly mitigated the large ice losses in West Antarctica and the Antarctic Peninsula \citep{Velicogna2020}.

There is an ongoing debate on the question of whether or not these recent mass gains might result from an intensification of the hydrological cycle under the current trend of anthropogenic global warming \citep[e.g.][]{Medley2019}, and how the latter impacted the observed increase in air temperature. This has led to an increased consciousness of the necessity to dissociate anthropogenic global warming from natural seasonal, annual, decadal or multidecadal variability related to atmospheric and oceanic processes driving surface temperature changes in Antarctica (CLIVASH2k, PAGES2k, \citealt{Jones2019,Rahaman2019}). Since the early 2000s, a large number of studies have addressed large-scale climate mode impact on Antarctic temperatures and how these interfere with anthropogenic global warming \citep{Thomas2013,Abram2014,Jones2019,Rahaman2019,Ejaz2021, Li2021}. These studies essentially rely on either Antarctic Station meteorological records \citep{Clem2013,Clem2015,Clem2016,Ekaykin2017,Jones2019} or on proxies from firn and ice core records (mainly water stable isotopes $\delta$D and $\delta^{18}$O, which are temperature proxies), extending the observational window beyond direct observations \citep{Bertler2006,Gregory2008,Divine2009,Naik2010,Thomas2013,Abram2014,Goodwin2016,Ekaykin2017,Rahaman2019,Ejaz2021}. 

The climate modes most often considered to explain changes in Antarctic temperature are the El Niño~/~Southern Oscillation (ENSO) and the Southern Annular Mode (SAM), also known as Antarctic Oscillation (AAO), but other climate patterns are also addressed, such as the Pacific Decadal Oscillation (PDO), the Interdecadal Pacific Oscillation (IPO), the Indian Ocean Dipole (IOD), and the Atlantic Multidecadal Oscillation (AMO) \citep{Eayrs2021,Fogt2022}. The methods used to link these large-scale climate modes to changes in Antarctic temperature have shown increasing sophistication with time, from simple correlations and anomaly composite studies \citep{Bertler2006,Gregory2008,Divine2009,Naik2010,Clem2013,Thomas2013,Abram2014,Goodwin2016,Clem2015,Clem2016,Ekaykin2017}, $\chi^2$ probability analysis \citep{Fogt2011}, to Principal Component Analysis (PCA) \citep{Ejaz2021}, power spectral analysis \citep{Divine2009,Naik2010,Rahaman2019,Ejaz2021} and wavelet analysis \citep{Gregory2008,Divine2009,Rahaman2019,Ejaz2021}.

Focus has often been placed on the Antarctic Peninsula and the West Antarctic Ice Sheet, given its significant warming in the last decades, with major interest in the teleconnection with tropical regions via ENSO, and its potential interaction with SAM, recognized as a major player in the control of Antarctic climate \citep{Fogt2020}. However, several studies have extended the research to other regions of Antarctica (Table~\ref{table1}): West Antarctica \citep{Thomas2013,Jones2019}, McMurdo Sound \citep{Bertler2006}, Princess Elizabeth Land \citep{Ekaykin2017,Crockart2021} and Dronning Maud Land, notably supported by recent acquisition of ice core data \citep{Divine2009,Naik2010,Ekaykin2017,Rahaman2019,Ejaz2021}.

\begin{sidewaystable*}[ht!]
\scriptsize
\caption{Studies linking isotope and temperature data from ice cores, weather stations and Hadley Centre Global Sea Ice and Sea Surface Temperature (HadISST) with climate indices in Antarctica. A cross indicates that a link between an index and an ice core is found in a specific study; if the time scale of influence is known, it is indicated instead of a cross. A potential link (not certain) is indicated in parenthesis. DML stands for `Dronning Maud Land'.}
\begin{tabular}{lcccccccc}
\toprule
\textbf{Reference} & \textbf{Location} & \textbf{Data source} & \textbf{ENSO} & \textbf{SAM} & \textbf{PDO} & \textbf{IPO} & \textbf{IOD} & \textbf{AMO} \\
\midrule
\citet{Bertler2006} & McMurdo Dry Valleys & Snow pits & Summer only & X & & & & \\
\citet{Gregory2008} & Southwest Antarctic Peninsula & Ice cores (ITASE, 2001-2,3,5) & More coastal 2001-5 & X & X & & & \\
\citet{Divine2009} & Coastal DML, West of Novolazarevskaya & 8 ice cores (including S800) & 2-4 yrs & >10 yrs & & & & \\
\citet{Naik2010} & DML, upstream Novolazarevskaya & Ice core IND25/B5 & (Maybe) & 4-10 yrs & & & & \\
\citet{Thomas2013} & West Antarctica & Ice core Ferrigno F10 & (Longer time scales) & 4-10 yrs & & & & \\
\citet{Clem2013} & Antarctic Peninsula & Weather stations & X & X & & & & \\
\citet{Abram2014} & Antarctic Peninsula (Northern tip) & James Ross ice core & X & X & & & & \\
\citet{Clem2015} & Antarctic Peninsula & Weather stations & X & & X & & & \\
\citet{Goodwin2016} & Antarctic Peninsula (North) & Bruce Plateau ice core (accumulation) & X & X & Multidec. & Multidec. & & \\
\citet{Clem2016} & Antarctic Peninsula & Weather stations & X & X & & & & \\
\citet{Ekaykin2017} & Princess Elizabeth Land & 6 ice cores and weather stations & & Interannual & & Interannual & >27 yrs \\
\citet{Jones2019} & West Antarctica & Weather stations and HadISST1 & & X & &  &  \\
 & and Antarctic Peninsula & & & & & & \\
\citet{Rahaman2019} & South Antarctic Peninsula / & Ice core stacks & 2-8 yrs & & 16-64 yrs & & & \\
 & West Peninsula vs. DML & & & & & & & \\
\citet{Crockart2021} & Mount Brown & Mount Brown ice core (salts) & X & & & & & \\
\citet{Ejaz2021} & DML, upstream Novolazarevskaya & IND33 ice core & X & 4-8 yrs & & & & \\
This study & DML, Princess Ragnhild Coast & FK17 & 4-5 and & 3-5 yrs & 2-3 and & (4-5 and & 2-5 yrs & \\
 & & & 7-10 yrs & & 5-7 yrs & 7-9 yrs) & & \\
This study & DML, Princess Ragnhild Coast & TIR18 & 2-5, 6-7 &  & 2-3 yrs & (3-4 yrs) & 5 yrs & 3-10 yrs \\
 & & & and 7-10 yrs & & & & & \\
\bottomrule
\end{tabular}
\label{table1}
\end{sidewaystable*}

How ENSO and SAM interact with Antarctic temperature is even more blurred by the fact that, as shown by \citet{Fogt2011} in their seminal study based on $\chi^2$ test probabilities on standardized monthly values of these two indices for the period 1957-2009, they are not independent, with a stronger correlation in spring-summer. These authors showed that, when ENSO and SAM are "in phase", the large-scale pressure fields between tropical regions and the Antarctic associated with SAM and ENSO strengthen, resulting in increasing teleconnection between the two regions. This has been confirmed by several ice core case studies (e.g. \citealt{Gregory2008,Naik2010,Ejaz2021}). On the other hand, during "out of phase" or neutral periods, the anomaly signatures of ENSO or SAM differ and weaken. Strong Rossby wave trains (teleconnections) of ENSO events are kept during "in phase" ENSO-SAM, while considerably weakened during "out of phase" ENSO-SAM.

\begin{figure}
\includegraphics[width=14cm]{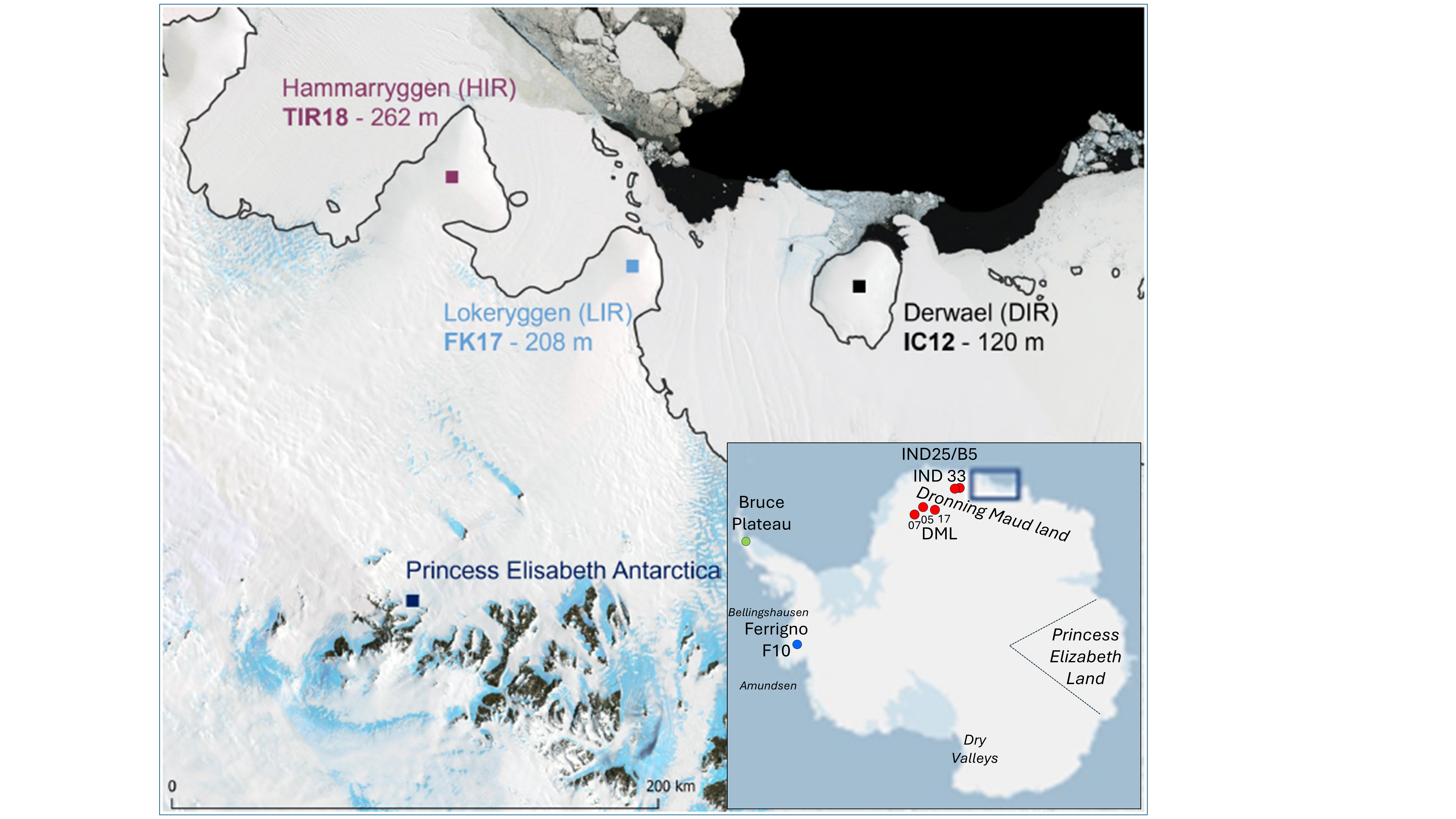}
\caption{Map of the study area with location of FK17 and TIR18 ice cores (this study) as well as IC12 \citep{Philippe2016} in Dronning Maud Land (more details in Sect.~\ref{data}). Ice core locations shown on inset map are from \citet{Ejaz2021} (red), \citet{Goodwin2016} (green) and \citet{Thomas2013} (blue).}
\label{map}
\end{figure}

In their recent review, \citet{Fogt2020} underline that there are both seasonal variations in the strength of the SAM-Antarctic temperature relationship and sign reversals of the relationship through time, particularly near the Antarctic Peninsula and parts of East Antarctica. Along the Antarctic Peninsula and western Dronning Maud Land, the connection is further complicated by the influence from ENSO, with anomalies during individual ENSO events changing significantly in time. For example, \citet{Naik2010} underline that when taking out El Niño and La Niña events from the 4-6 year band pass record of $\delta^{18}$O and SAM in the IND25/B5 ice core (Dronning Maud Land, Figure~\ref{map}), a significant negative correlation emerges, suggesting  that indeed ENSO events weaken the SAM-temperature relationship (i.e., colder during positive SAM events). These anomalies differ not only in time, but also regionally in space between western Peninsula, North-East Peninsula or western Dronning Maud Land (e.g. \citealt{Clem2013,Ejaz2021}). Local topographic effects can also impact the relationship between atmospheric indices and temperature, such as the strengthened Foehn effect across the Antarctic Peninsula during positive SAM, inducing warming on the eastern side \citep{Orr2004,Marshall2006,Goodwin2016}, a process eventually strengthened by anthropogenic forcing.

SAM and ENSO impacts on the temperature records have also been isolated in other parts of Antarctica (Table~\ref{table1}). \citet{Bertler2006} used reconstructed summer temperatures from snow pits in the McMurdo Dry Valleys (Figure~\ref{map}) to show that these are controlled by the relative intensity of summer Southern Oscillation Index (SOI) and SAM. These authors suggest that the connection resides in ENSO impact on the Amundsen Sea Low intensity and its control on katabatic wind intensity. In Princess Elizabeth Land (Figure~\ref{map}), \citet{Ekaykin2017} use inland ice core data and coastal weather stations to calibrate the $\delta$D-temperature relationship, which they compare to the SAM index, showing significant correlation both at the annual and multidecadal scales during the post-industrial era.

Several studies have used power spectral analysis, the Multi Taper Method (MTM) and the wavelet spectral analysis to extract potential periodicity in both climate modes and temperature records \citep{Gregory2008,Divine2009,Naik2010,Rahaman2019,Ejaz2021}. Cross-wavelet analysis has also been used to detect potential co-periodicity between climate modes, or between climate modes and temperature proxies \citep{Divine2009,Rahaman2019}. In the West Dronning Maud Land area (Figure~\ref{map}), \citet{Divine2009} detected four modes of $\delta^{18}$O variability in the S100 ice core using MTM (quasi biannual, quadriennal, decadal and bi-decadal), suggesting correlation to ENSO at shorter time scales (bi-annual and quadriennal) and to SAM at longer time scales (decadal and bi-decadal). \citet{Naik2010} used Power Spectrum Redfit 3.5 for the $\delta^{18}$O record of the IND25/B5 DML ice core (upstream of the Novolazarevskaya Station, Figure~\ref{map}) and detected significant peaks at 4 years (they associate them to SAM) and other peaks at 6 years (SOI) and 10 years (SAM). Applying an 8-12 year band pass to SAM and $\delta^{18}$O datasets, these authors detected a significant correlation ($R$~=~-0.72), concluding that SAM dominates the isotopic variability on decadal time scales, with intermittent influence of ENSO. More recently, \citet{Ejaz2021} described the long-term evolution of the reconstructed temperature record in the IND33 ice core from the same area in Dronning Maud Land (Figure~\ref{map}). The study detects a cooling trend from 1842 to 1907 and a warming trend in the 1942-2019 period. Coherent with the study of \citet{Abram2014}, they note that the recent warming is synchronous with an increase in ENSO events and in strong anti-phase with SAM. This behaviour is confirmed by ERA5 reanalysis showing that La Niña events lead to warming in the whole Peninsula and the Weddell Sea and cooling in Dronning Maud Land, while El Niño results in warming in the South-West Peninsula, cooling in the North-East Peninsula and warming in Dronning Maud Land. The authors suggest that the increase of the SAM index under the anthropogenic forcing of the last decades implies negative pressure anomalies over 60$^\circ$S, strengthening the impact of El Niño events in the Dronning Maud Land area (despite the interannual anti-phase SOI-SAM relationship described above), and hence the warming.

The detection of multi-decadal periodicity in some of the $\delta^{18}$O records has led some authors to look at synergies with longer time-scale climate indices (Table~\ref{table1}), such as PDO and IPO for the Pacific domain \citep{Gregory2008,Clem2015,Goodwin2016,Rahaman2019} or IOD for the Indian Ocean domain \citep{Ekaykin2017}. As proposed by \citet{Goodwin2016}, the common signature found for PDO and IPO in the $\delta^{18}$O record of the Bruce Plateau ice core (tip of Antarctic Peninsula) suggests they describe a Pacific wide ocean-atmosphere climate variability, affecting the position of the South Pacific Convergence Zone (SPCZ) and therefore tropical-Antarctic teleconnections. \citet{Ekaykin2017} show a strong correlation of their ice core stack from Princess Elizabeth Land reconstructed temperature with IOD at a 27-year band-pass. They reckon that the mechanism is presently unknown, although probably linked to cyclonic activity.

We focused in the above sections on climate modes as potential drivers for Antarctic $\delta^{18}$O-temperature records. However, it is widely recognized that other factors should be considered to explain water isotope variability in ice cores, such as sea ice extent, seasonal bias in the Antarctic precipitation or post-depositional redistribution at a given location, as well as anthropogenic global warming (e.g. \citealt{Divine2009,Kuttel2012,Thomas2013,Goodwin2016,Ekaykin2017,Ejaz2021}).

Following earlier work \citep{Bromwich1983,Noone2004}, several authors demonstrated that the ice core $\delta^{18}$O record shows a significant relationship with sea ice concentration, extent or area in the neighbouring sectors, raising the opportunity to use it as a proxy for past sea ice extent \citep{Kuttel2012,Thomas2013,Holloway2016,Holloway2017,Thomas2019,Ejaz2021}. The mechanism behind the sea ice - $\delta^{18}$O relationship would actually involve two competing effects: a reduction in the maximum winter sea ice extent would increase the proportion of winter precipitation, hence depleting $\delta^{18}$O at the ice core sites; however, because of the reduced source-to-site distance for atmospheric vapor, the $\delta^{18}$O would be enriched due to lesser fractionation. The joint overall effect is for $\delta^{18}$O to increase as sea ice is reduced \citep{Holloway2016}. \citet{Kuttel2012} demonstrate this relationship for 8 West Antarctic ice cores under the influence of the sea ice extent variability in the Amundsen Sea, while \citet{Thomas2013} used a similar approach for the Ferrigno F10 ice core (at the western root of the Antarctic Peninsula), with negative correlation to winter sea ice extent and area in the Bellingshausen and Amundsen Seas. Both studies, however, underline that it is not always trivial to dissociate the imprint of the sea ice driver from the one of the condensation temperature at the site. In a recent study in West Dronning Maud Land ice cores (coastal IND33 vs. continental DML 05-07-17), \citet{Ejaz2021} nicely addressed this challenge. Using PCA, they showed that the two first principal components PC1 and PC2 grouped ca. 60$\%$ of the total variance of their $\delta^{18}$O records, with the dominant PC1 related to temperature and PC2 related to sea ice concentration in the West Indian sector. They further showed that PC1 concerns the coastal IND33 core, while PC2 dominates the inland cores. Using the $\delta^{18}$O PC2-sea ice concentration relationship, these authors then reconstructed sea ice concentration from 1800 onwards. Finally, on larger time scales, \citet{Holloway2016,Holloway2017} demonstrated, using a water isotopes-enabled General Circulation Model (GCM), that it is a drastic retreat of the winter sea ice (43-67$\%$) that was primarily responsible for the warm $\delta^{18}$O peak observed in continental deep Antarctic ice cores during the Eemian (128ky), rather than the total collapse of the Western Antarctic ice Sheet.

It is interesting to note that all the studies addressing the impact of sea ice extent on ice core $\delta^{18}$O underline the connection to the larger-scale atmospheric circulation patterns (meridional and zonal winds), the latter being modulated by atmospheric indices such as ENSO, SAM or multidecadal PDO. Correlation between sea ice concentration and atmospheric indices might therefore underly correlation between $\delta^{18}$O-derived temperatures and atmospheric indices,  raising the issue of direct causality.

This paper investigates the sources of low-frequency $\delta^{18}$O variability in two coastal ice cores located in Dronning Maud Land (Antarctica), close to the Belgian Princess Elisabeth Station, for which measurements start at the end of the 18th century, focusing on various climate indices and sea ice area as potential drivers. The ice core data as well as the climate indices and the sea ice observations used here are described in Sect.~\ref{data}. We apply three different approaches, encompassing spectral and novel causal analyses, in order to facilitate the identification of potential sources of variability, which are described in Sect.~\ref{methods}. The spectral analyses allow initial isolation of  the key spectral peaks, while the causal analyses allow for getting the proper directional influences between the climate and sea ice indices and the temporal variability of the ice cores. To our best knowledge, this is the first time such a combination is applied, which allows for getting robust conclusions on the sources of $\delta^{18}$O variability in Dronning Maud Land. Our results are discussed in Sect.~\ref{results} and our conclusions are presented in Sect.~\ref{conclusions}.

\section{Data}
\label{data}

\begin{figure}
\includegraphics[width=14cm]{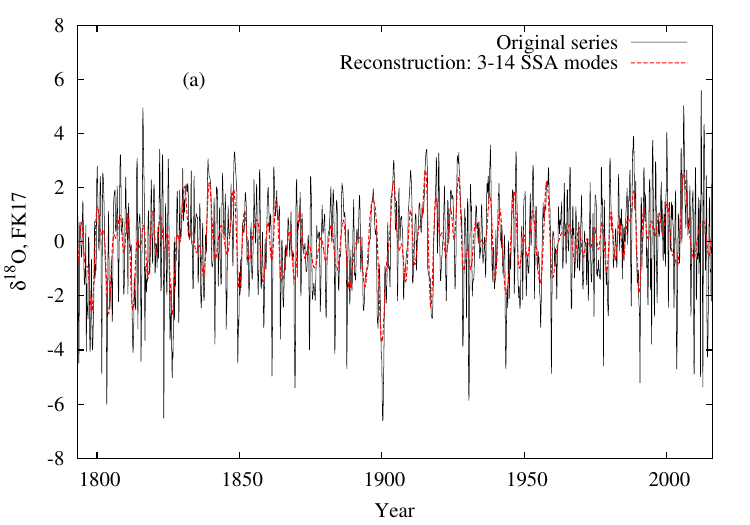}
\includegraphics[width=14cm]{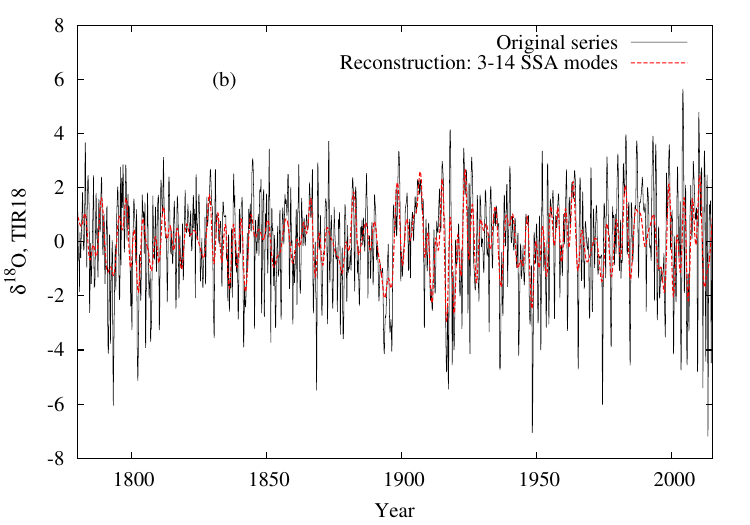}
\caption{Temporal evolution of the $\delta^{18}$O in the ice cores at sites FK17 (a) and TIR18 (b). The black curve is the raw standardized time series, while the red curve is the reconstruction of the low-frequency variability based on a selection of 3-14 modes obtained by a Singular Spectrum Analysis (see text for details).}
\label{Series}
\end{figure}

Our two study sites are located on the Princess Ragnhild Coast, in East Dronning Maud Land. Two ice cores, FK17 (208~m) and TIR18 (262~m), were drilled at the crest of two adjacent ice rises (Lokeryggen Ice Rise LIR, 70.53648$^\circ$S, 24.07036$^\circ$E, and Hammarryggen Ice Rise, 70.49960$^\circ$S, 21.88017$^\circ$E), ca. 100 km apart. Both ice rises are actually ice promontories, with inland connection (Figure~\ref{map}). Only the top 120 meters of each ice core were analyzed at high resolution (5~cm) for density, water stable isotopes ($\delta^{18}$O and $\delta$D), major ions and DC electrical conductivity (ECM). Details of the field work and of the dating process are presented in \citet{Wauthy2024}, together with profiles of selected environmental proxies. A main outcome of that study is that both environmental proxies and surface mass balance show contrasting behaviors, suggesting strong spatial and temporal variability at the regional scale. The location of the ice cores provides an ideal setting to document further the discussion initiated in the previous studies on the potential impact of large-scale climate modes and sea ice extent on the $\delta^{18}$O-temperature record of coastal ice cores, stretching the area of investigation further to the East in Dronning Maud Land (Sect.~\ref{intro} and Figure~\ref{map}). In this study, we will focus on the two $\delta^{18}$O profiles shown in Figure \ref{Series}. $\delta^{18}$O was measured using a PICARRO L 2130-i cavity-ring down spectrometer (CRDS). The long-term reproducibility (i.e., standard deviation of internal standards) is 0.02 $\%$ \citep{Wauthy2024}. Figure \ref{Series} displays the two standardized (shift of the mean to 0 and divided by the standard deviation) monthly time series (continuous black curves) of the $\delta^{18}$O in the ice cores at sites FK17 and TIR18. 

Monthly sea ice area (SIA) from 1979 to 2010 is obtained from the National Snow and Ice Data Center \citep{Meier2021}. SIA is calculated as the sum of all grid cells with sea ice concentration greater than 0~$\%$, multiplied by their areas, and is expressed in km$^2$. The data is then organized in bins of 5$^\circ$ of longitude. In our analysis, we consider three sectors, namely 0-20$^\circ$E, 20-40$^\circ$E and 40-55$^\circ$E, due to their relative proximity to the FK17 and TIR18 ice cores. In order to compute the rate of information transfer (Sect.~\ref{information_transfer}), we need to get approximately stationary time series, thus we remove the linear trend and seasonality from SIA.

We use seven different regional climate indices characterizing the atmosphere and ocean (Table \ref{table2}), which are known to exert some influence on the Antarctic climate \citep{Eayrs2021,Fogt2022}, and thus potentially on $\delta^{18}$O at the two sites analyzed here. Time series of these indices were retrieved from the Physical Sciences Laboratory (PSL) of the National Oceanic and Atmospheric Administration (NOAA; \url{https://psl.noaa.gov/data/climateindices/list/}, last access: 2 February 2023). We use monthly values from January 1979 to December 2010 (384 months), and we remove the linear trend.

The Southern Annular Mode (SAM), or Antarctic Oscillation (AAO), is constructed by projecting the 700~hPa geopotential height anomalies poleward of 20$^\circ$S onto the leading Empirical Orthogonal Function (EOF; using monthly mean 700~hPa height anomalies from 1979 to 2000). It is computed based on the National Centers for Environmental Prediction / National Center for Atmospheric Research (NCEP/NCAR) reanalysis. The SAM is the dominant pattern of natural variability in the Southern Hemisphere outside the tropics. When the SAM is in its positive phase, the westerly wind belt that drives the Antarctic Circumpolar Current intensifies and contracts towards Antarctica. When the SAM is negative, the westerly wind belt weakens and moves towards the equator.

The Niño3.4 index, hereafter referred to as `Niño', tracks the oceanic part of ENSO. It is based on standardized sea-surface temperature (SST) anomalies (using version 5 of the NOAA Extended Reconstructed SST [ERSST]) averaged over the eastern tropical Pacific (5$^\circ$S-5$^\circ$N; 170-120$^\circ$W). The Niño3.4 index is in its warm phase when SST anomaly exceeds 0.5$^\circ$C (El Niño), and it is in its cold phase when SST anomaly is below -0.5$^\circ$C (La Niña).

The Southern Oscillation Index (SOI) tracks the atmospheric part of ENSO. It is computed as the standardized difference in sea-level pressure between Tahiti (central tropical Pacific) and Darwin (Australia, western tropical Pacific), based on NCEP/NCAR reanalysis. The positive (negative) phase of the SOI represents above-normal (below-normal) air pressure at Tahiti and below-normal (above-normal) air pressure at Darwin, and corresponds to La Niña (El Niño).
	
The Pacific Decadal Oscillation (PDO) is obtained by projecting the Pacific SST anomalies from ERSST v5 dataset onto the dominant EOF from 20$^\circ$N to 60$^\circ$N. The PDO is positive when SST is anomalously cold in the interior North Pacific and warm along the Pacific Coast. The PDO is negative when the climate anomaly patterns are reversed.

The Tripole Index (TPI) for the Interdecadal Pacific Oscillation (IPO) is based on the difference in SST anomalies between the average over the central equatorial Pacific (10$^\circ$S–10$^\circ$N, 170$^\circ$E–90$^\circ$W) and the average over the Northwest (25$^\circ$N–45$^\circ$N, 140$^\circ$E–145$^\circ$W) and Southwest Pacific (50$^\circ$S–15$^\circ$S, 150$^\circ$E–160$^\circ$W). We use the unfiltered Hadley Centre Global Sea Ice and Sea Surface Temperature version 2.1 (HadISST2.1) data, and the methodology for computing TPI is explained in \cite{Henley2015}.
    
The Atlantic Multidecadal Oscillation (AMO) is computed based on version 2 of the \cite{Kaplan1998} extended SST gridded dataset (which uses UK Met Office SST data) averaged over the North Atlantic (0-70$^\circ$N; unsmoothed time series) and following the procedure described in \cite{Enfield2001}.

Finally, we use the Dipole Mode Index (DMI), which characterizes the intensity of the Indian Ocean Dipole (IOD). It is computed as the difference in SST anomalies between the western equatorial Indian Ocean (50$^\circ$E-70$^\circ$E, 10$^\circ$S-10$^\circ$N) and the southeastern equatorial Indian Ocean (90$^\circ$E-110$^\circ$E and 10$^\circ$S-0$^\circ$N), based on HadISST1.1  data. When DMI is positive (negative), SST is larger than average in the western (eastern) Indian Ocean.

\begin{sidewaystable*}[ht!]
\footnotesize
\caption{Atmospheric and oceanic indices used in this study.}
\begin{tabular}{llcccc}
\toprule
\textbf{Acronym} & \textbf{Full name} & \textbf{Variables} & \textbf{Location} & \multicolumn{2}{c}{\textbf{Sign}} \\
\midrule
 &  &  &  & \textbf{Positive} & \textbf{Negative} \\
\midrule
SAM & Southern Annular Mode & 700 hPa & South of 20$^\circ$S & Westerly wind belt intensifies & Westerly wind belt moves \\
 & & geopotential height &  & and contracts towards Antarctica & towards the equator \\
\midrule
Niño & Niño3.4 index & Sea-surface temperature & 5$^\circ$S-5$^\circ$N; & El Niño: & La Niña: \\
 &  & (SST) & 170-120$^\circ$W & Warm phase & Cold phase \\
\midrule
SOI & Southern Oscillation Index & Sea-level pressure & Tahiti (central tropical & La Niña: & El Niño: \\
 &  & (SLP)  & Pacific) - Darwin & Higher pressure at Tahiti & Lower pressure at Tahiti \\
 &  &  & (western tropical Pacific) & Lower pressure at Darwin & Higher pressure at Darwin \\
\midrule
PDO & Pacific Decadal Oscillation & SST & 20-60$^\circ$N & Cold North Pacific & Warm North Pacific \\
 &  &  &  & Warm Pacific Coast & Cold Pacific Coast \\
\midrule
TPI & Tripole Index & SST & \textbf{1:} 10$^\circ$S-10$^\circ$N; &  & \\
(IPO) & (Interdecadal Pacific Oscillation) &  & 170$^\circ$E-90$^\circ$W &  & \\
 &  &  & \textbf{2:} 25$^\circ$N-45$^\circ$N; & Warm tropical Pacific & Cool tropical Pacific \\
 &  &  & 140$^\circ$E-145$^\circ$W & Cool northern Pacific & Warm northern Pacific \\
 &  &  & \textbf{3:} 50$^\circ$S-15$^\circ$S; &  & \\
 &  &  & 150$^\circ$E-160$^\circ$W &  & \\
\midrule
AMO & Atlantic Multidecadal Oscillation & SST & 0-70$^\circ$N & Warm North Atlantic & Cold North Atlantic \\
\midrule
DMI & Dipole Mode Index & SST & \textbf{1:} 10$^\circ$S-10$^\circ$N; &  & \\
(IOD) & (Indian Ocean Dipole) &  & 50$^\circ$E-70$^\circ$E & Warm western & Cold western \\
 &  &  & \textbf{2:} 10$^\circ$S-0$^\circ$N; & Indian Ocean &  Indian Ocean \\
 &  &  & 90$^\circ$E-110$^\circ$E &  & \\
 \bottomrule
\end{tabular}
\label{table2}
\end{sidewaystable*}

\section{Methods}
\label{methods}

Extracting signals from the noise is a long standing problem \citep{Smith2020}, which is even more complicated when dealing with multi-scaling chaotic dynamics as illustrated for instance in \cite{Vannitsem2021}. A question that arises in this context is to know what is the source of low-frequency variability, either from intrinsic dynamics or from an external forcing of the system at hand. In the current work, the main purpose is to isolate the external forcing that could potentially be at the origin of the low-frequency variability in the ice core time series. 

In order to extract information on the sources of $\delta^{18}$O variability in the ice cores at sites FK17 and TIR18 related to the climate and sea ice indices described in Sects.~\ref{intro} and \ref{data}, several methods are used. The Singular Spectrum Analysis (SSA), which allows for extracting the dominant modes of variability in each series, will be used in order to elegantly the dominant seasonal signal and the high frequency modes (e.g. \citealt{Ghil2002}). The power spectrum of the original and reconstructed (with SSA) signals will be extracted using the Multi Taper Method that provides robust estimates of the energy at each frequency (e.g. \citealt{Ghil2002}). In parallel, the causal influences between the low-frequency variability and the external climate indices and SIA is assessed through the estimation of the rate of information transfer proposed in \cite{Liang2021}, which will properly inform about the potential source of variability by extracting the directional influence of some indices on the $\delta^{18}$O. These three methods are now described in more details.  

\subsection{Multi Taper Method (MTM)}

Spectral analysis is a primary tool to extract the key frequencies at the origin of the variability of dynamical processes. Many approaches have been proposed along the years as discussed in details in the review of \cite{Ghil2002}. One method that has gained popularity is the Multi Taper Method (MTM), which allows for reducing the variance of the spectral amplitudes by averaging over a set of independent estimates of the power spectrum. This is done by multiplying the time series by orthogonal tapers chosen to minimize the spectral leakage due to the finite size of the data \citep{Percival1993, Ghil2002}. The specific choice of tapers, and hence the number of spectra on which the averaging is performed, depends on the length of the time series, and experiments with a few hundreds of data points indicate that three separate estimates is reasonable as suggested in \cite{Ghil2002}. This method is implemented in a software developed along the years by Ghil and collaborators \citep{Vautard1989,Vautard1992,Groth2015}, and called the SSA-MTM toolkit that can be downloaded at \url{http://research.atmos.ucla.edu/tcd/ssa/}.

Note that in the software, there is a possibility to test the emergence of frequencies against a colored (correlated) noise  or white (uncorrelated) noise, see \cite{Ghil2002} for a detailed discussion. The colored noise estimates are based on classical first order autoregressive models or continuous Ornstein-Uhlenbeck processes that show spectral slope at high frequencies of $-2$. As in our data set this spectral slope is close to $-3$, the colored noise approach does not fit well to the spectra of the current data. We therefore do not rely on such a test, but rather focus on the assumption of local white noise in the spectral domain. In the analysis below the local window covers a spectral range of $0.1$ year$^{-1}$. Following \cite{Thompson1990}, the variability of the estimates is made and an F-test applied. Confidence intervals of a pure local white noise can be proposed, and if the actual frequencies are larger than these levels, the local white noise hypothesis is rejected. The significance level chosen here to reject the null hypothesis is fixed to 1$\%$, reflected in Fig.~\ref{MTM-raw} by the 99$\%$ confidence interval of the white noise hypothesis. Note that in Fig. \ref{MTM-raw}, the median of the estimates is also provided for reference.

\subsection{Singular Spectrum Analysis (SSA)}

The Singular Spectrum Analysis (SSA) displays strong similarities with the principal component analysis (PCA), which is meant to select key spatial modes, the main difference being that its purpose is to extract the temporal dominant modes. To construct these modes, the time series, $X(i)$ with $i=1,..., N$, is embedded into a phase space of dimension, say $M$, whose coordinates are the successive values in the time series (e.g. \citealt{Broomhead1986, Fraedrich1986, Ghil2002}). The evolution in this phase space is then obtained by sliding the M-window in time. This operation can be visualized as

\begin{equation}
\begin{pmatrix}
X(1) & X(2) & X(3) & ... & X(N') \\
X(2) & X(3) & X(4) & ... & X(N'+1) \\
....      \\
X(M-1) & X(M) & X(M+1) & ... & X(N'+M-1) \\
X(M) & X(M+1) & X(M+2) & ... & X(N'+M) \\
\end{pmatrix}
\end{equation}
where $N'=N-M$. The covariance $M \times M$ matrix of these vectors can then be computed, and the eigenvalues and eigenvectors can be extracted. These eigenmodes will characterize the dominant modes present within the $M$-window selected. Selecting a few of these dominant modes will then allow for filtering out the high-frequency modes and the annual cycle that are disregarded in the current work. The window $M$ is usually fixed to $1/10$ of the length of the time series in order to have enough statistics for the covariance matrix. In our study, it is fixed to $268$ months, i.e. a bit more than $20$ years, and the filtering is performed for the two $\delta^{18}$O time series (FK17 and TIR18). The details of the method are provided in \cite{Ghil2002}.

The MTM analysis can also be performed on these filtered series. This will allow to isolate the key frequencies associated with the dominant modes of variability, and to check whether the spectral peaks found in the original time series are robust when removing the lower modes of variability usually associated with rapid noisy processes.

\subsection{Rate of information transfer}
\label{information_transfer}

The last decades have seen the development of techniques to disentangle the directional dependencies between different observables, going beyond the classical symmetric correlation analysis. Correlation provides information on the co-variability between two variables, but not on the causality between them: a simple example is the correlation between the monthly temperature at some location in the Northern Hemisphere and the monthly temperature at another location in the Southern Hemisphere; both are correlated because of the seasonal cycle, but they are not necessarily causal to each other. To clarify potential influences, causal methods indicating directions of influences should be used \citep[e.g.][]{Runge2019}. The Liang's method is providing precisely such a tool based on the evolution of the Shannon information entropy of a target variable, and isolating the flow of information/uncertainty from another variable of the system to the target. For a Gaussian process, this reduces to evaluating the impact of a variable of the system on the variance of the target one. 

In a series of works, Liang developed a methodology based on the estimation of the changes of theinformation entropy of the system, leading to a closed expression for the rate of information transfer  between different observables \citep{Liang2005,Liang2014a,Liang2014b,Liang2016,Liang2021}. This quantity was originally deduced in a quite general context of nonlinear stochastic systems, and subsequently deduced for linear stochastic systems. The advantage of the latter extension is to allow for an estimation based on observational data as discussed in \cite{Liang2014a,Liang2014b,Liang2021} and applied in various climate contexts \citep{Vannitsem2019,Vannitsem2022, Docquier2022, Docquier2023}.    

Consider $S$ time series, $X_i$, $i=1,...,S$,
all having $N$ equidistant data points $X_i(n)$, $n=1,2,...,N$ and for which one can estimate a temporal derivative
such that
  \begin{eqnarray*}
  \dot X_i(n) = \frac{X_i(n+1) - X_i(n)} {\Delta t}
  \end{eqnarray*}
where $\Delta t$ is the time step.
Let $C_{ij}$ be the sample covariance between $X_i$ and $X_j$, and $C_{i,dj}$ the sample covariance between $X_i$ and $\dot X_j$. It has been shown the estimator of the rate of information transfer from, say, $X_j$ to $X_i$, is
	\begin{eqnarray}	 \label{eq:Tji_hat}
	\hat T_{j \to i} = \frac 1 {\det {\bf C}} \cdot 
		       \sum_{k=1}^S \Delta_{jk} C_{k,di}
			\cdot \frac {C_{ij}} {C_{ii}},
	\end{eqnarray}
where $\Delta_{jk}$ are the cofactors of the covariance matrix $ {\bf{C}} =
(C_{ij})$, and $\det \bf{C}$ is the determinant of $\bf{C}$.

Note that generically causation implies correlation, but correlation does not imply causation. This feature is also revealed by the mathematical expressions above, when dealing with only two time series as illustrated in \cite{Liang2014a}. So the test that should be made to evaluate the causal dependencies is to check whether the rate of information transfer is significantly different from 0. Different approaches can be used, and here we will use a bootstrap method with replacement \citep{Efron1993}, in a similar way as in \cite{Vannitsem2019}. 

Note that the results that will be presented in the following are normalized in such a way that the influences of all observables on one specific target (here $\delta^{18}$O) are put on the same ground. It will give a percentage of influence of the other observables on the target. The normalization factor in the multivariate case is as in \cite{Liang2021}:
\begin{equation*}
Z_i =  \sum_{k=1}^{S} |T_{k \rightarrow i} | + \left|  \frac{d H_i^{noise}}{d t} \right|
\end{equation*}
where $ \left| \frac{d H_i^{noise}}{d t} \right|$ is the noise contribution.

The relative transfer of information from $X_j$ to $X_i$ is then given by,
\begin{equation*}
\tau_{j \to i} = \frac{T_{ j \rightarrow i}}{Z_i}     
\end{equation*}

This quantity will be displayed for the set of observables presented in Sect.~\ref{data} that may influence the $\delta^{18}$O. 

Note that in order to select the proper time scales at which the dependencies act, moving averages are used as in \cite{Vannitsem2022}. This approach is acting as a low-pass filter that selects the  time scales at which processes interact. Note that this selection is complex as illustrated in the systematic analysis of temporal averages on the properties of chaotic systems \citep[e.g.][]{Nicolis1995, Vannitsem1995}. Furthermore, this approach does not reduce much the number of data points in the time series, allowing to get enough statistics for computing the rate of information transfer.  

\section{Results and discussion}
\label{results}

In this section, the spectral analyses of the emerging low-frequency variability of $\delta^{18}$O are first performed. This is followed by an analysis of the causality between the climate indices and sea ice area on the one hand and the two $\delta^{18}$O time series on the other hand. Finally, a discussion is provided synthesizing the potential links found using the causal analysis and comparing to previous results from the literature.

\subsection{Low-frequency variability}

\begin{figure}
\includegraphics[width=14cm]{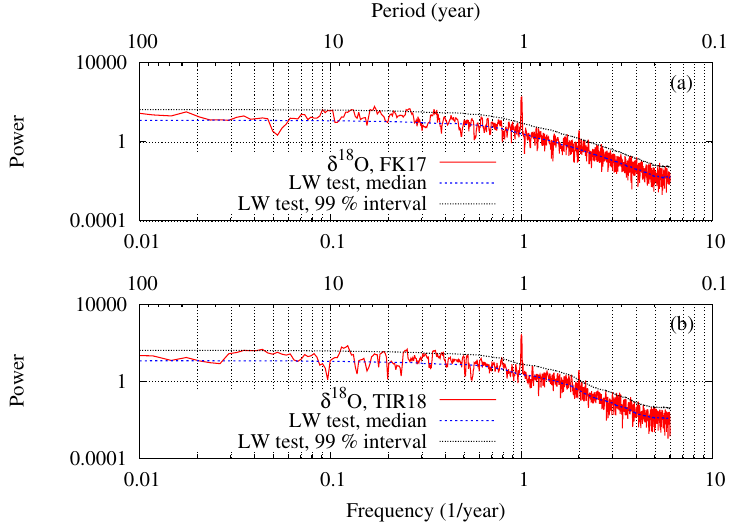}
\caption{Power spectrum of the original $\delta^{18}$O in the ice cores at sites FK17 (a) and TIR18 (b), as obtained with the Multi Taper Method (MTM) developed in the Package of \cite{Ghil2002} (continuous red curve). The dashed and dotted lines represent the median and the upper bound of the 99$\%$ confidence interval of the local white noise modeling assumption. LW in the panels refers to Local White hypothesis testing.}
\label{MTM-raw}
\end{figure}

The power spectra of the two original $\delta^{18}$O time series computed with the MTM approach are displayed in Fig.~\ref{MTM-raw} and summarized in Table~\ref{table3} (row 3). These spectra display a classical shape with a peak at the yearly time scale associated with the seasonality, a rapid power law-like decrease at high frequencies (with a slope around -3), and a few peaks at low frequencies. For FK17, one can qualitatively isolate low-frequency peaks passing the 1~$\%$ significance level at periods of approximately $3, 4, 6, 8$ and $10-12$ years. For TIR18, peaks are found at about $3, 4, 8-9$ and $22-28$ years. Interestingly the two spectra at such large temporal scales seem to partly differ, suggesting that different mechanisms might be at play in producing such low frequency variability.

\begin{table*}[ht!]
\caption{Periods identified by the MTM-SSA analysis and time scales of influence identified by the causal method for each climate driver.}
\begin{tabular}{lcccccc}
\toprule
 & \multicolumn{2}{c}{\textbf{Climate driver}} & \multicolumn{2}{c}{\textbf{FK17}} & \multicolumn{2}{c}{\textbf{TIR18}} \\
\cmidrule(l{3pt}r{3pt}){2-3}
\cmidrule(l{3pt}r{3pt}){4-5}
\cmidrule(l{3pt}r{3pt}){6-7}
 & \textbf{Name} & \textbf{MTM period} & \textbf{Original data} & \textbf{SSA filtered data} & \textbf{Original data} & \textbf{SSA filtered data} \\
\midrule
\textbf{MTM-SSA} &  &  & 3, 4, 6, 8, 10-12 & 3, 4, 6, 8, 10 & 3, 4, 8-9, 22-28 & 3, 5, 7 \\
\midrule
\textbf{Causal} & SAM & 2-3, 4-5 & & 3-5 & & \\
\cline{2-7}
\textbf{method} & Niño & 4-5 & 4-5, 7-10 & & & 2-3 \\
\cline{2-7}
 & SOI & 2, 4, 5-20 & 4, 10 & 8-10 & 3-5, 6-7, 9-10 & 4, 7-10 \\
\cline{2-7}
 & PDO & 1, 2, 4-18 & 2-3, 5-6 & 2-3, 5-7 & 4-5 & 4-5 \\
\cline{2-7}
 & TPI & 4-5 & 4-5, 7-9 & & 3-4 & \\
\cline{2-7}
 & AMO & multi-decadal & & 10 & 3-9 & 4-6, 9-10 \\
\cline{2-7}
 & DMI & 3-5 & & 2-5 & & 5 \\
\cline{2-7}
 & SIA 0-20$^\circ$E & 3-5 & & 5-6 & & \\
\cline{2-7}
 & SIA 20-40$^\circ$E & 3-6 & & 0-3 & & 6 \\
\cline{2-7}
 & SIA 40-55$^\circ$E & 5-7 & & & 5-7 & 5-8 \\
\bottomrule
\end{tabular}
\label{table3}
\end{table*}

The relevance of these peaks with respect to the possible influence of climate drivers is of course the central question of the current work, but before going into the causal analysis, one can isolate the low-frequency variability from the broadband high-frequency variability and the seasonal cycle. To this aim, we use the SSA to select the dominant modes acting at such scales. 

For FK17, the SSA spectrum is showing a pair of high energetic modes at the seasonal cycle for 18.5 $\%$ of the variance (not shown). The 12 next modes show a low frequency variability and account for about 36 $\%$ of variance. The following modes show a more complex behavior with a mixing of low and high frequencies. We therefore decide to restrict ourselves to these 3 to 14 modes for the reconstruction of the series. As shown in the upper panel of Fig.~\ref{Series} (dotted red curve), the evolution of the reconstruction follows closely the original series, while removing the seasonal cycle and high frequency variability. 

For TIR18, the two first modes, also accounting for the seasonal cycle, represent 22.2 $\%$ of the variance. The modes from 3 to 14 are also selected, similar to FK17, and account for 33 $\%$ of the variance. This choice is made based on the visual analysis of the temporal variability of the SSA modes and to have a comparable set of modes as for FK17. The result is shown in the lower panel of Fig. \ref{Series}, which indeed provides a good representation of the signal low-frequency variability at TIR18. 

\begin{figure}
\includegraphics[width=14cm]{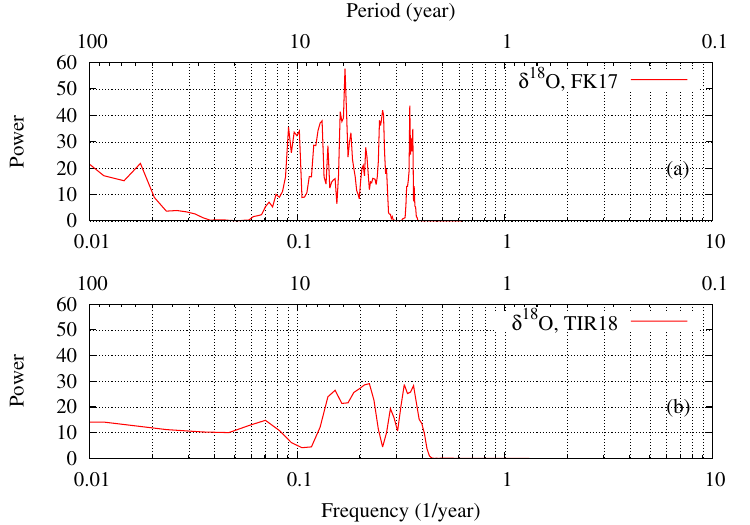}
\caption{Power spectrum for the SSA reconstruction of the ice core $\delta^{18}$O at sites FK17 (a) and TIR18 (b), as obtained with the Multi Taper Method.}
\label{MTM-recons}
\end{figure}

Let us now compute the power spectra of these reconstructed time series. Figure~\ref{MTM-recons} displays the two spectra in a linear-log scale in order to isolate the key peaks in the spectrum, also summarized in Table~\ref{table3} (row 3). There is a global consistency of low frequency peaks between the original data sets and the dominant modes supporting robustness. For FK17, the peaks isolated in the original spectrum now look more pronounced and well separated at $3, 4, 6, 8$ and $10$ years. For TIR18, the peaks are broader, with less maximal power than at FK17 and centred on $3, 5$ and $7$ years. There is thus a common peak at $3$ years, and two-three peaks at longer times scales.

\subsection{Sources of low-frequency variability and links to climate and sea ice}

In order to disentangle the potential sources of low-frequency variability in the two time series, we now compute the rate of information transfer between $\delta^{18}$O and the set of selected indices presented in Sect.~\ref{data}. Note that in the current causality analysis, the data used only cover the period from 1979-2010 for which reliable estimates are available for all indices considered (satellite observations of sea ice concentration start in 1979 and TPI data are available until 2010). This limitation implies that one cannot explore a too large averaging time window. We limit ourselves here to a maximum 10-year time window. 

\begin{figure}
\includegraphics[width=18cm]{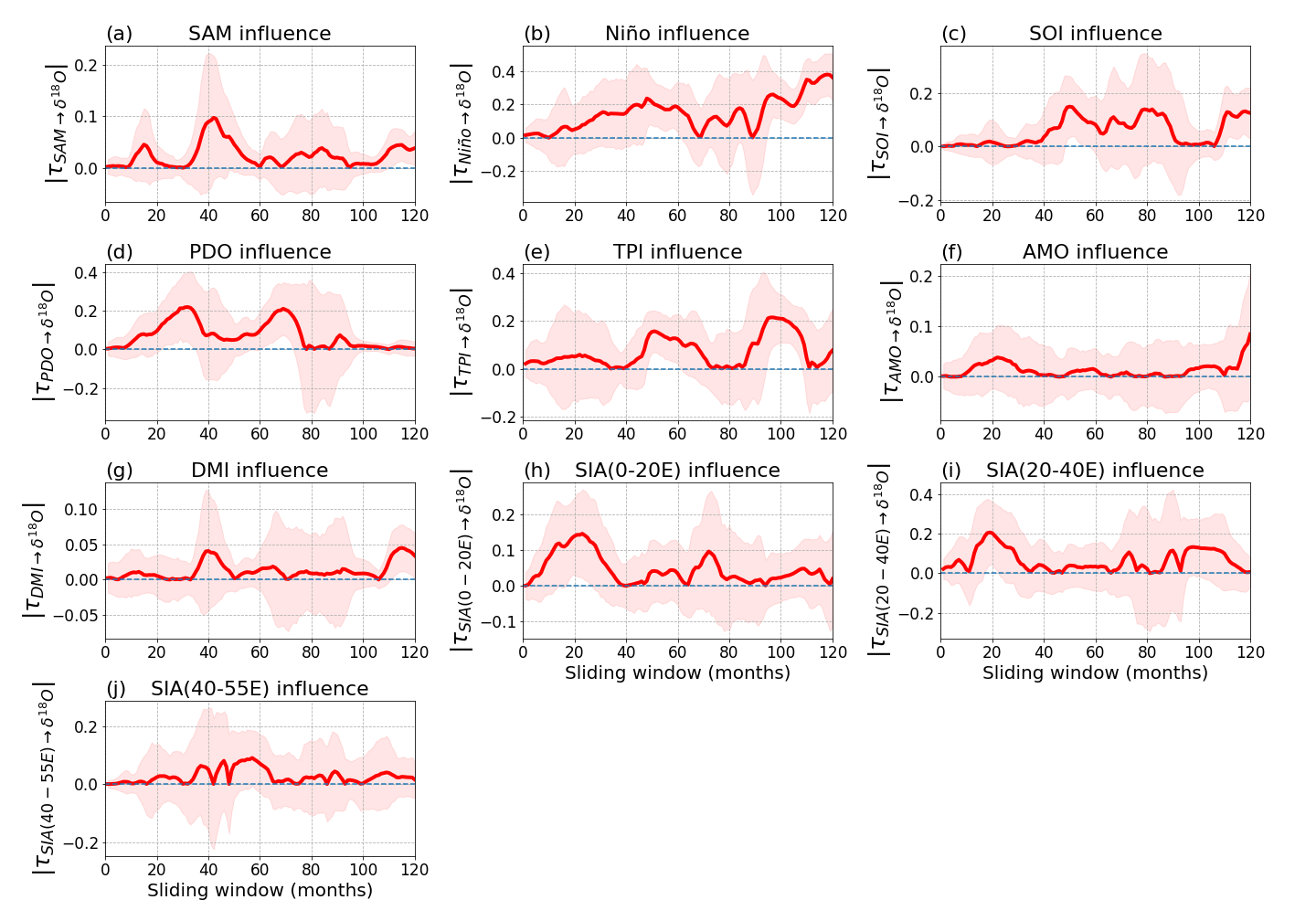}
\caption{Rate of information transfer (absolute value) as a function of the sliding window, from the key climate indices to the original $\delta^{18}$O at FK17. The red shading represents the 99~$\%$ confidence interval; if this interval does not contain the 0 value, the information transfer is statistically significant.}
\label{Liang-raw-FK17}
\end{figure}

\begin{figure}
\includegraphics[width=18cm]{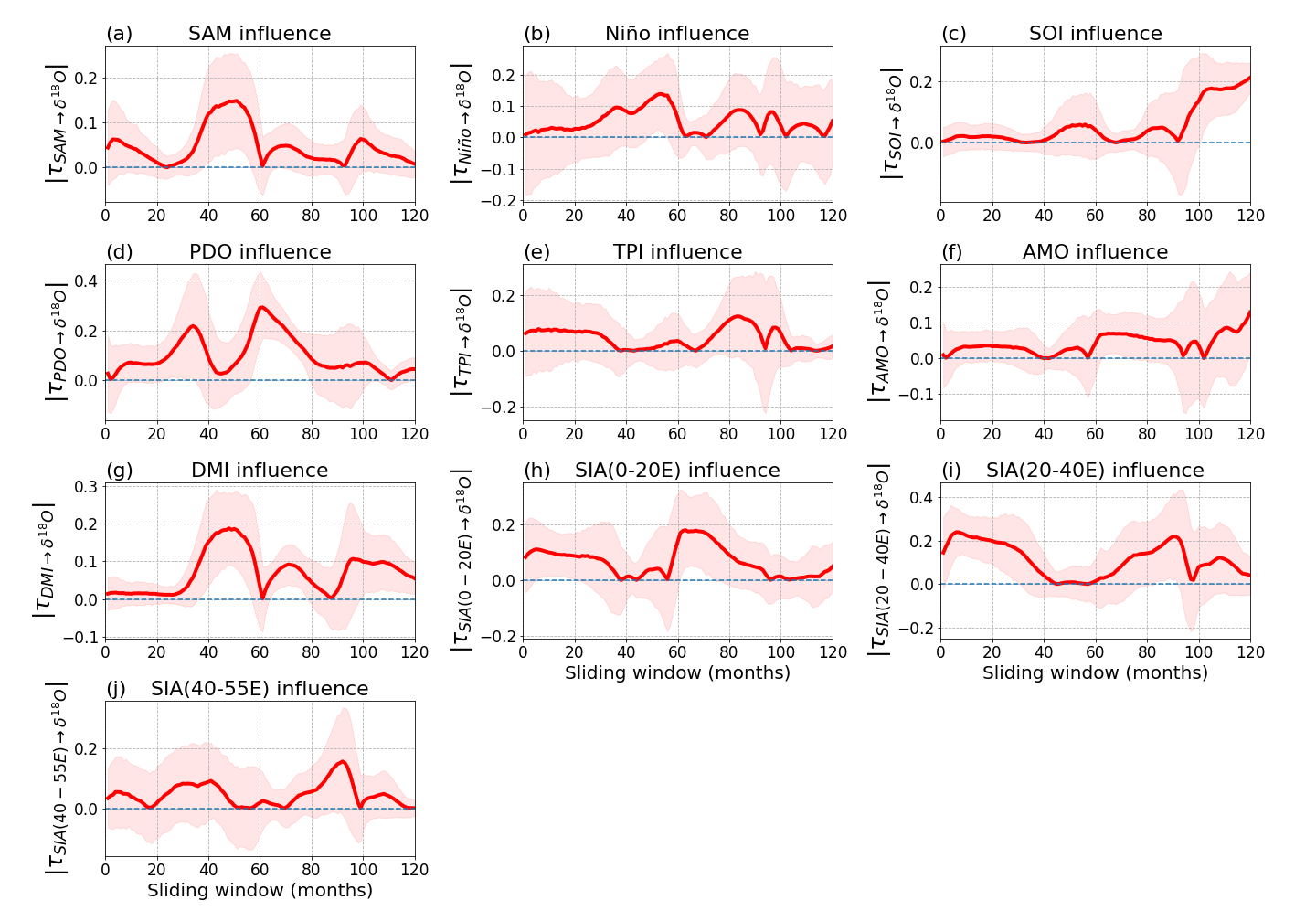}
\caption{Rate of information transfer (absolute value) as a function of the sliding window, from the key climate indices to the reconstructed $\delta^{18}$O at FK17 based on SSA. The red shading represents the 99~$\%$ confidence interval; if this interval does not contain the 0 value, the information transfer is statistically significant.}
\label{Liang-recons-FK17}
\end{figure}

\begin{figure}
\includegraphics[width=18cm]{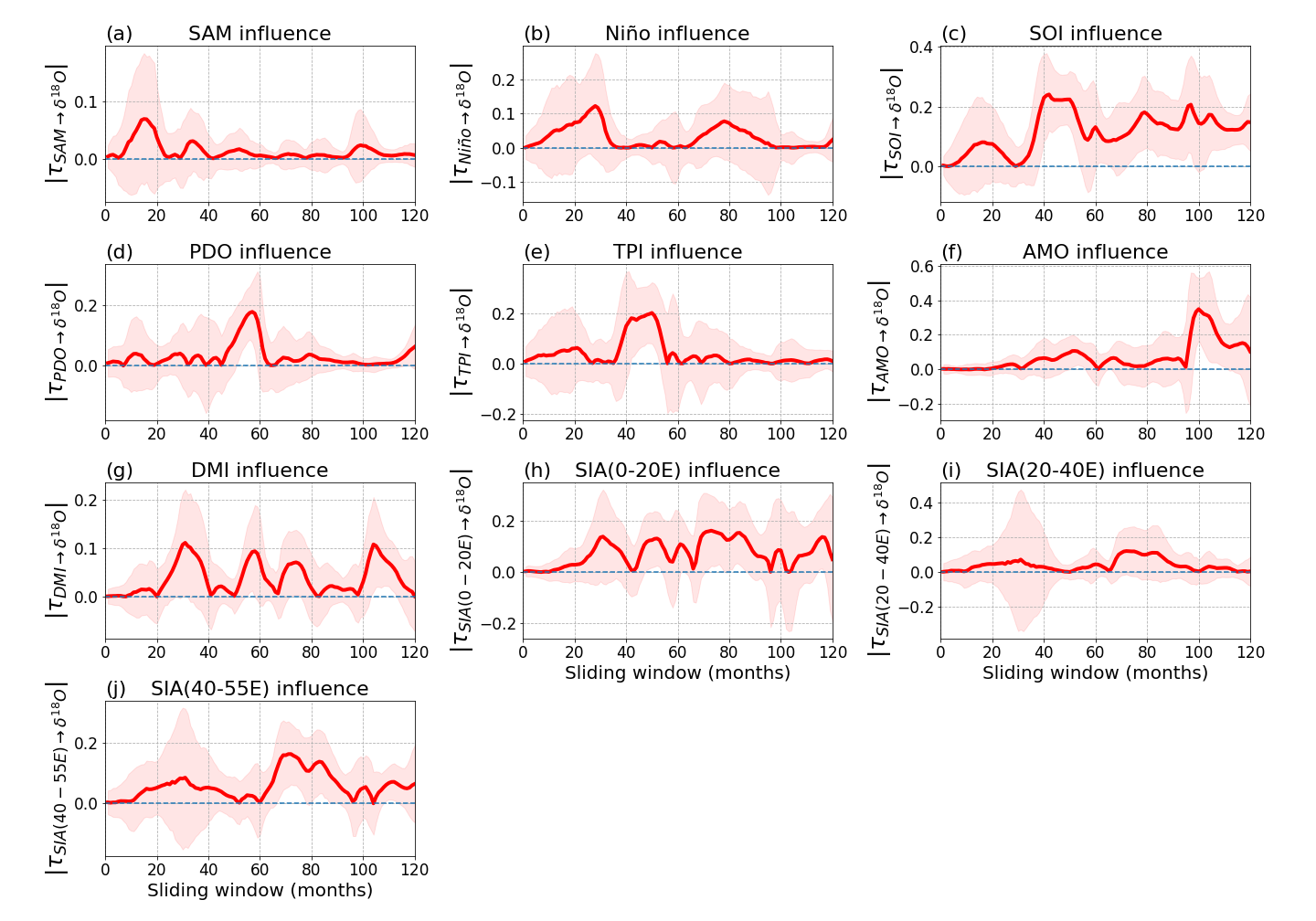}
\caption{Rate of information transfer (absolute value) as a function of the sliding window, from the key climate indices to the original $\delta^{18}$O at TIR18. The red shading represents the 99~$\%$ confidence interval; if this interval does not contain the 0 value, the information transfer is statistically significant.}
\label{Liang-raw-TIR18}
\end{figure}

\begin{figure}
\includegraphics[width=18cm]{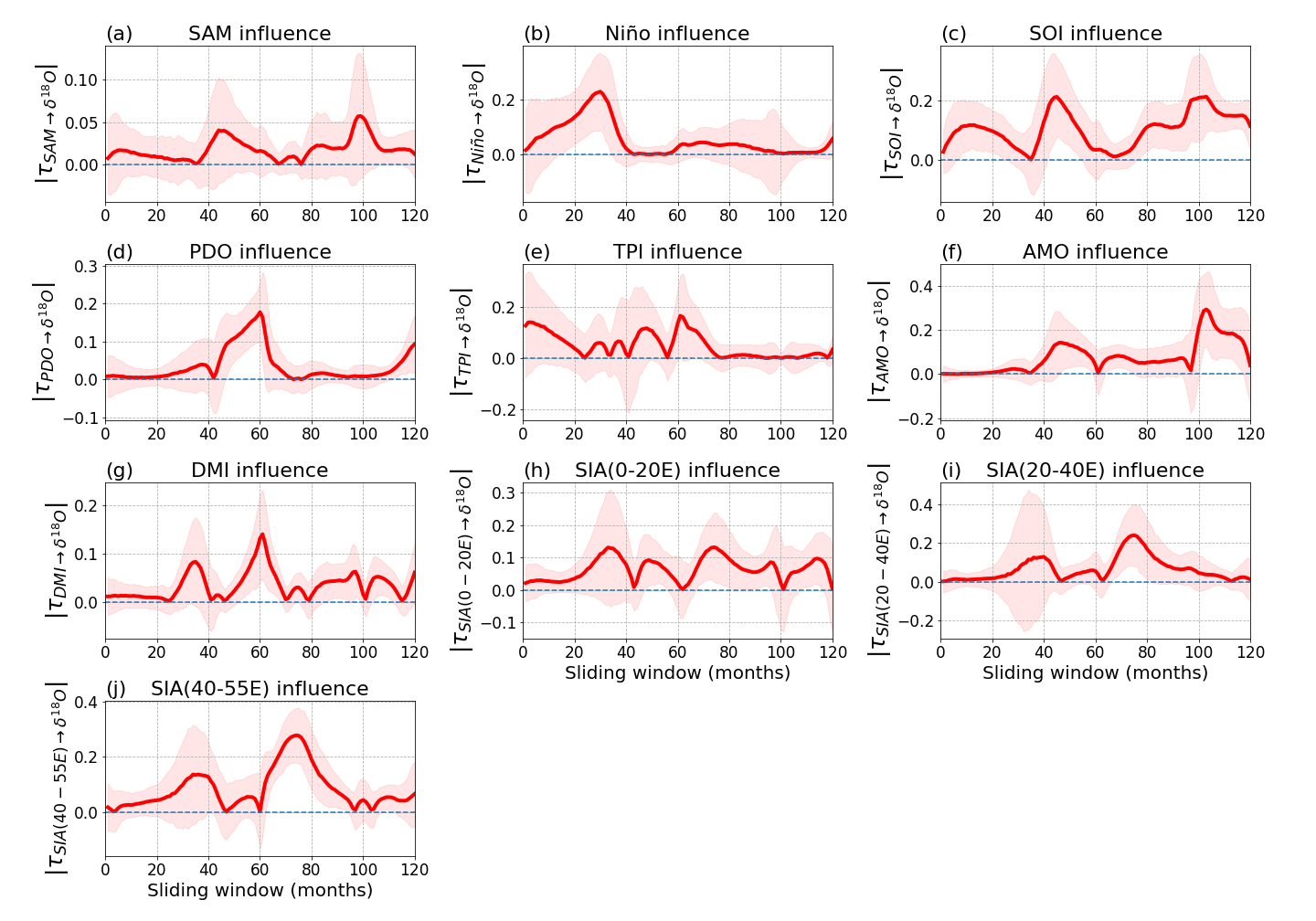}
\caption{Rate of information transfer (absolute value) as a function of the sliding window, from the key climate indices to the reconstructed $\delta^{18}$O at TIR18 based on SSA. The red shading represents the 99~$\%$ confidence interval; if this interval does not contain the 0 value, the information transfer is statistically significant.}
\label{Liang-recons-TIR18}
\end{figure}

\begin{figure}
\includegraphics[width=18cm]{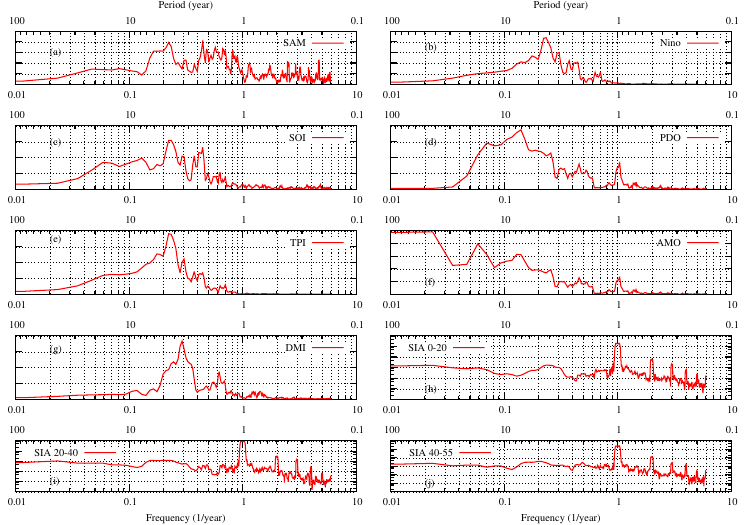}
\caption{Multi Taper Method applied to the set of climate and sea ice indices. }
\label{MTM-indices}
\end{figure}

Figure~\ref{Liang-raw-FK17} displays the rate of information transfer for the original time series of $\delta^{18}$O at FK17 as a function of the moving time window. The shading band represents the uncertainty of the estimation at the 1~$\%$ significance level. If the zero value line lies within the 99~$\%$ interval, we consider that there is no significant dependence. The large uncertainty is typical of the method when applied to short datasets as it also needs to estimate time derivatives that are usually affected by larger errors. Figure~\ref{Liang-recons-FK17} reproduces the same analysis with the FK17 reconstructed time series, for which only dominant low-frequency variability extracted from the SSA is considered  in order to check the robustness of the analysis. Figures~\ref{Liang-raw-TIR18} and \ref{Liang-recons-TIR18} show the same analyses as Figs.~\ref{Liang-raw-FK17}-\ref{Liang-recons-FK17}, respectively, for the TIR18 data set. Figure~\ref{MTM-indices} summarizes the results from the MTM analysis, for the climate and sea ice indices that potentially contribute to the ice core $\delta^{18}$O low-frequency variability, as detailed in the following paragraphs. Table~\ref{table3} synthesizes the salient features from the whole set of figures and results to support the discussion below.

Interestingly, a large number of significant causal influences emerge from Table~\ref{table3}. As for the power spectra analysis, there is a relatively good coherence between the whole data set (`Original data' in Table~\ref{table3}) and the dominant modes (`SSA filtered data' in Table~\ref{table3}) at both locations. There are however a few exceptions. In some cases, a strong dependence is present for the original data set, which disappears in the SSA analysis (i.e., the TPI index at both locations, the Niño index for longer time windows, the SOI index for shorter time windows at FK17, the SOI index at intermediate time windows, and the SIA in the 20-40$^\circ$E sector at TIR18). This suggests a complex interplay between the different indices at different time scales. There are also a few cases where a strong signal appears in the SSA filtered analysis which did not show up for the original data set due to the filtering of the fast SSA modes (Niño short-term, AMO intermediate-term and SIA 20-40$^\circ$E sector at TIR18, SAM, AMO decadal and DMI at FK17).

We will now discuss in more details the significance of the causal influences between our $\delta^{18}$O records and the chosen climate and sea ice indices, including the coherence between the main MTM periods of those indices and the  significant causal influences (Figure~\ref{MTM-indices} and Table~\ref{table3}). In general, there is no perfect match between the spectra of the forcing and the forced signal when the forcing is chaotic, as already pointed out by \cite{Vannitsem2021}, but the comparison can still be used if expressed in terms of periodicity ranges, as done in the following.

SAM has overall a weak influence on our isotopic signature, with only a short-term impact of $3-5$ years in the SSA analysis at FK17. ENSO, on the contrary, is well present, both on shorter ($2-5$ years) and longer ($7-10$ years) periodicities at both locations. However, atmospheric (SOI) and oceanic (Niño) indices show a different behaviour, with SOI responding at both locations and Niño more sensitive at FK17. The Niño short-term response and both SOI short-term and long-term responses fit with the main periodicities of the indices.

The PDO short to intermediate-term periodicities ($1$ to $7$ years) also emerge at both locations. TPI is only significant (also at short to intermediate time scales) in the original data set. AMO also shows a significant causal influence at short to intermediate time scales, but only clearly at TIR18. Finally, DMI is significant at its appropriate periodicity in the SSA analysis at both locations.

The impact of SIA is generally present at both locations, with however some differences related to which sectors affect most $\delta^{18}$O at the ice core location. FK17 is influenced by SIA at sectors 0-20$^\circ$E and 20-40$^\circ$E, while TIR18 is poorly sensitive to sector 0-20$^\circ$E, with more impact from sectors 20-40$^\circ$E and 40-55$^\circ$E.

To summarize, our analysis puts forward the following salient causal influences:
\begin{itemize}
    \item In the recognized fluctuating balance of SAM-ENSO influence on coastal Antarctic $\delta^{18}$O, it appears that our very coastal ice cores are dominated by ENSO, both on the shorter and longer terms
    \item While SOI (atmospheric derived) affects similarly both locations, FK17 shows more connections to the Niño index (oceanic derived)
    \item The SAM influence is only present at FK17, with a short-term periodicity
    \item PDO and IOD (DMI) all show significant causal influences on $\delta^{18}$O at both locations, at short to intermediate term periodicities, corresponding to their power spectra
    \item The influence of AMO on the $\delta^{18}$O at TIR18 is present for time windows of 8-9 years. Such a peak is present in the power spectrum (Fig. \ref{MTM-indices}f) but it is difficult to isolate it from the rest of the spectrum as higher power are present at even larger scales, i.e. this could be considered as a fluctuation in the scaling of the power spectrum. This reduces our confidence in the existence of this link and more detailed analyses of this specific influence with longer climate series would be needed.  
    \item IOP (TPI) has also an impact, but not in the dominant SSA modes
    \item SIA also has a general impact, with a potentially more eastern influence at TIR18
\end{itemize}

The ENSO influence has been reported in all other ice cores from the western DML region \citep{Divine2009, Naik2010, Rahaman2019, Ejaz2021} within the $2-8$ years periodicity range (Table~\ref{table1}). This is coherent with the ranges we find at FK17 and TIR18, although distributed in discrete windows ($2-5$, $6-7$ and $7-10$ years, Table~\ref{table3}). SAM also has a clear influence on the $\delta^{18}$O signal in the above-mentioned studies, for periods larger than 4 years. The SAM influence is less present in our study, only at FK17 and only in the SSA analysis, and at the lowest end of the range ($3-5$ years). 

The other climate indices are more difficult to compare, since they have been less often considered in previous work. PDO was detected in the earlier study of \cite{Gregory2008} in the ITASE cores of the South-West Antarctic Peninsula, and also considered in later studies based on Antarctic Peninsula weather stations \citep{Clem2016} or ice cores \citep[Bruce Plateau; ][]{Goodwin2016}. \cite{Rahaman2019} invoke its impact on long time scales ($16-64$ years) in the DML IND33 ice core. The PDO influence is clearly present in our study, at both locations in the $2-3$ and $5-7$ year intervals, coherent with the broad $4-18$ range (peaking at $7$ years) of its power spectrum (Table~\ref{table3}, Fig.~\ref{MTM-indices}). The IPO signature is also present in our cores, but only in the original data set and not in the dominant mode. It has also been invoked at the Bruce Plateau ice core \citep[][Antarctic Peninsula]{Goodwin2016} and in Princess Elizabeth Land, much further East \citep{Ekaykin2017}. Princess Elizabeth Land also showed an impact of IOD at long time scales (> $27$ years). The IOD influence is also potentially present in both our cores, but at shorter times scales ($2-5$ years). It does not show up in the other studies from the Dronning Maud Land area, either because these are located further to the West, or because it may not have been investigated in these studies. Finally, we found a significant causal influence of AMO on the TIR18 ice core at $3-10$ years, not detected anywhere else.

As mentioned previously, SIA has also been considered as a driver for  $\delta^{18}$O records in ice cores. We have considered the monthly SIA record in the longitudinal sector of our two ice core locations (20$^\circ$-40 $^\circ$E), but also in the surrounding sectors immediately to the West (0$^\circ$-20$^\circ$E) and to the East (40$^\circ$-55$^\circ$E). All three sectors clearly show a causal influence of SIA on $\delta^{18}$O at both locations, but with a predominance of the western and central sectors at FK17 and of the central and eastern sectors at TIR18. This could explain part of the contrast in proxy signatures between the two locations \citep{Wauthy2024}, as also suggested by preliminary air mass backtracking studies in the area, showing a TIR18 air mass source extending more easterly and continentally influenced.

\section{Conclusions}
\label{conclusions}

A methodology to disentangle the external source of low-frequency variability in Dronning Maud land ice cores has been proposed by first analyzing the spectral properties of the series with a multi-taper method (MTM) and the singular spectrum analysis (SSA), and second, by using a causality analysis based on tools from the information theory, which is a novel approach in this field.

The spectral analysis of the two $\delta^{18}$O time series (FK17 and TIR18) recorded during the last 250 years in the Dronning Maud Land reveals a dynamics with a dominant low-frequency variability from $2$ to $15$ years. Beside this common feature between the two sites, the shape of the spectra at these scales is quite different, with the emergence of multiple sharp peaks for FK17, while a broadband peak appears for TIR18. 

The analysis of the origin of this low-frequency variability from external sources reveals a large spectrum of causal influences, underlining the complexity of the task. ENSO, SAM, PDO, DMI and SIA show important causal influences on FK17, while for TIR18, the main influences are from ENSO, PDO, DMI, AMO and SIA. It is the first time that the influence of the IOD (DMI) is brought forward in DML ice cores, which is expected given the more easterly location of these cores compared to previous studies. Note however that differences in the amplitude of the common influences at both locations are present, together with the time windows at which they emerge. All together, these might reveal differences in the source of air masses that feed the two ice cores, and therefore observed differences in proxy behaviour. The explanations for these differences are still not clear and need further investigations. Furthermore, this study has indeed focused on the $\delta^{18}$O signature of the cores, but it could easily be generalized to the other proxies described in \cite{Wauthy2024}, e.g. deuterium excess, surface mass balance, non-sea salt sulfates, methanesulfonic acid, which also show strong local contrasts.

It is interesting that the two independent approaches, i.e. the MTM/SSA spectral analyses and causal method provide relatively similar periodicities (Table~\ref{table3}). However, the methods also have their limitations. First, given the length of the data set considered, the combined causal and spectral analyses cannot explore the domain of long-term periodicities ($>10$ years), which might be of relevance when looking at the original $\delta^{18}$O and other proxy records of the ice cores in the area investigated (FK17 and TIR18 - \cite{Wauthy2024}, but also IC12 - \cite{Philippe2016}). Future analyses on longer available climate indices could help disentangling the impact of recent anthropogenic warming from natural multidecadal climatic modes. Second, while the causal method allows to detect influences and remove spurious dependencies based on correlations, it does not indicate the relative importance of the mechanisms involved in the actual build-up of the $\delta^{18}$O signature (e.g. condensation temperature at the site, distillation process on the way from source to location, seasonality of precipitation). In this regard, a pathway for future research would be to apply the same approach to a more restricted data set sorted out by PCA techniques, as was done for correlations in the past (\citealt{Ejaz2021}, see Sect.~ \ref{intro}), or the Multi-variate Singular Spectrum Analysis \citep{Ghil2002}. Finally, the dependence analysis based on the Liang's approach assumes the linearity of the overall dynamical influence of the external sources on the $\delta^{18}$O. This leaves the possibility of nonlinear influences that could be missed in our analyses. A recent work has been performed in that direction by designing a fully nonlinear approach \citep{Pires2024}, which has been recently tested on idealized models \citep{Vannitsem2024}. This method promises to be highly relevant for real world case studies, and it will be explored in the future.

\section{acknowledgements}
This collaboration started under the support of the Belgian Federal Science Policy Office (BELSPO) in the context of the Mass2Ant project (contract no. BR/165/A2/Mass2Ant). 
David Docquier and Stéphane Vannitsem were also partly supported by ROADMAP (Role of ocean dynamics and Ocean-Atmosphere
interactions in Driving cliMAte variations and future Projections of impact-relevant extreme events; https://jpi-climate.eu/project/roadmap/),
a coordinated JPI-Climate/JPI-Oceans project, and the funding from BELSPO under contract B2/20E/P1/ROADMAP. David Docquier is currently funded by BELSPO under the RESIST project (contract no. RT/23/RESIST).
Sarah Wauthy benefited from an FRS-FNRS research fellowship, and was supported by the Mass2Ant project (BELSPO) and the PARAMOUR project (supported by the FNRS and the FWO under the Excellence of Science (EOS) program; grant no. O0100718F, EOS ID no. 30454083).
The FK17 and TIR18 $\delta^{18}$O datasets with other records including ion concentration, electrical conductivity measurements, density, surface mass balance, and age models, are combined in a file titled “Physico-chemical properties of the top 120 m of two ice cores in Dronning Maud Land (East Antarctica)”. This file is available on Zenodo under the Creative Commons Attribution 4.0 International Public License \citep{Wauthy2023}.


\bibliographystyle{wileyqj}
\bibliography{biblio}

\end{document}